\documentclass{jpsj2}
\newcommand{\eg}{{e.g., }}
\newcommand{\ie}{{i.e., }}
\newcommand{\vecB}{{\bf B}}
\newcommand{\vecf}{{\bf f}}
\newcommand{\vecF}{{\bf F}}
\newcommand{\vecr}{{\bf r}}
\newcommand{\vecsigma}{{\boldsymbol\sigma}}
\newcommand{\vecv}{{\bf v}}
\newcommand{\xhat}{{\hat{\bf x}}}

\newcommand{\zhat}{{\hat{\bf z}}}

\title{Hydrodynamic Interaction in Confined Geometries}

\author{Haim \textsc{Diamant}\thanks{E-mail: hdiamant@tau.ac.il}}

\inst{School of Chemistry, Tel Aviv University, Tel Aviv 69978, Israel}

\abst{ This article gives an overview of recent theoretical and
experimental findings concerning the hydrodynamic interaction between
liquid-embedded particles in various confined geometries. A simple
unifying description emerges, which accounts for the various findings
based on the effect of confinement on conserved fields of the
embedding liquid. It shows, in particular, that the hydrodynamic
interaction under confinement remains long-ranged, decaying
algebraically with inter-particle distance, except for the case of
confinement in a rigid linear channel.  }

\kword{colloids, suspensions, microfluidics, response functions, pair diffusion,
pair mobility, drag, effective viscosity, porous media}

\begin{document}
\maketitle

\section{Introduction}
%=====================
\label{sec_intro}

Particles moving through a fluid affect each other's velocity through
the flow that their motions cause. These medium-induced, hydrodynamic
interactions play a crucial role in the dynamics of all particulate
liquids, such as colloid suspensions \cite{Russel} and polymer
solutions \cite{DoiEdwards}, and have been thoroughly studied.
In the past several years there has been significant progress in
clarifying the effects of spatial confinement on hydrodynamic
interactions. This research has been driven by new techniques for the
fabrication of microfluidic channels
\cite{microfluid} and for the tracking and manipulation of individual
particles \cite{tweezer,video,Bechinger}. The findings highlight the
dramatic effects that confinement on the scale of the size of
particles has on their hydrodynamic interaction. It can change the
sign of the interaction, its decay with distance, and its
concentration dependence
\cite{prl02,prl04,jpcm052}. 
Under conditions of driven flow it may also lead to a new type of
density waves \cite{Tsevi1,Tsevi2}. 

This article provides an overview of recent developments concerning
the hydrodynamic interaction in confined geometries. Rather than
summarizing various technical results, we attempt to present an
intuitive unifying description, which derives from the effects of
confining boundaries on the conserved fields of the fluid. To this end
it is helpful to begin by recalling, in the following section, a few
fundamentals concerning the hydrodynamic interaction in an unconfined
liquid. We then proceed in \S\ref{sec_q2D} to show how these basic
considerations are modified in three examples of confined,
quasi-two-dimensional (q2D) systems. Section \ref{sec_q1D} addresses
confinement in quasi-one-dimensional (q1D) channels, and in
\S\ref{sec_porous} we comment on hydrodynamic interactions in liquids
embedded in three-dimensional (3D) solid matrices, such as gels and
porous media. Finally, in \S\ref{sec_discuss} we discuss further
implications and open issues.

\section{General Considerations}
%===============================
\label{sec_3D}

The dynamics of a fluid can be coarse-grained into continuous
equations for its conserved fields \cite{LL}. In an isotropic fluid
these fields are the local densities of mass, momentum, and energy. If
the heat conductivity is much larger than the kinematic viscosity
$\nu$, thermal relaxation will be much faster than that of momentum,
and the temperature can be assumed uniform. If the sound velocity $c$
is much larger than $\nu/r$, where $r$ is the length scale under
consideration, sound (compressive) modes will be much faster than
transverse-momentum (shear) ones, and the mass density $\rho$ can be
assumed uniform. Finally, if the scales of length $a$ and velocity $v$
of the particles moving through the fluid are sufficiently small to
yield a negligible Reynolds number, $va/\nu\ll 1$, the inertial terms
in the (Navier-Stokes) equation for the momentum density can be
neglected. Although this combination of conditions may seem
restrictive, it actually holds in a broad range of circumstances
relevant to particulate liquids \cite{Happel}. In such a case the
steady-state flow satisfies the equations,
\begin{align}
  -&\nabla p + \eta\nabla^2 \vecv + \vecf = 0,
 \label{Stokes}\\
  &\nabla\cdot\vecv = 0,
 \label{incompress}
\end{align}
where $\vecv(\vecr)$ is the fluid velocity, $p(\vecr)$ its pressure,
$\vecf(\vecr)$ the applied force density, and $\eta=\rho\nu$ the
dynamic viscosity. Equations (\ref{Stokes}) and (\ref{incompress})
reflect, respectively, the conservation of momentum and mass in an
isothermal, incompressible liquid at zero Reynolds number.

The problem of finding the pair hydrodynamic interaction between
particle 1 at the origin and particle 2 at $\vecr$ amounts to solving
eqs.\ (\ref{Stokes}) and (\ref{incompress}) with $\vecf=0$, given that
the particles translate with velocities $\vecv^1$ and $\vecv^2$, and
subject to appropriate boundary conditions (\eg no slip) at the
surfaces of the particles and at the system boundaries. One can
subsequently calculate the forces $\vecF^1$ and $\vecF^2$ acting on
the particles, and from their linear dependence on the prescribed
velocities establish a pair mobility tensor,
\begin{equation}
  v^\alpha_i = B^{\alpha\beta}_{ij}(\vecr) F^\beta_j,\ \ 
  \alpha,\beta=1,2\ \ 
  i,j=x,y,z.
\end{equation}
(Summation over repeated indices is implied throughout the article.)
In particular, $B^{21}_{ij}(\vecr)$ gives the velocity of a force-free
particle 2 due to a force acting on particle 1, thus characterizing
the pair coupling. If we define the $x$ axis along $\vecr$, the
diagonal terms of $\vecB^{21}$ correspond to three coupling
``polarizations''---a longitudinal coupling and two transverse ones,
\begin{equation}
\begin{split}
  &B^{\rm c}_{\rm L}(r) = B^{21}_{xx}(r\xhat),\\
  &B^{\rm c}_{\rm T1}(r)= B^{21}_{yy}(r\xhat),\ \ 
  B^{\rm c}_{\rm T2}(r)= B^{21}_{zz}(r\xhat).
\end{split}
\label{Bc}
\end{equation}
For an unbounded, isotropic system these are also the eigenvalues of
$\vecB^{21}$ (\ie off-diagonal terms vanish), and the two transverse
coefficients are equal.
In the overdamped limit under consideration the Einstein relation
safely holds.  Thus, the coupling mobility coefficients are simply
related to coupling diffusion coefficients, $D^{\rm c}_{\rm
L,T}=k_{\rm B}T B^{\rm c}_{\rm L,T}$, $k_{\rm B}T$ being the thermal
energy. The coupling diffusion coefficients, in turn, can be directly
measured by tracking the correlated Brownian motion of particle pairs.

In the case of no-slip boundary conditions at the particle surfaces,
and in the limit of large inter-particle distance $r$ compared to the
particle size (radius) $a$, the procedure described above is
significantly simplified. To leading order in $a/r$ the force
distribution over the surface of particle 1 may be replaced by a force
monopole, $\vecf(\vecr)=\vecF^1\delta(\vecr)$, and the velocity of
particle 2 may be assumed equal to the would-be velocity of the liquid
at its position if it were absent, $\vecv^2=\vecv(\vecr)$. Thus, the
coupling mobility is equal in this limit to the velocity Green's
function of eqs.\ (\ref{Stokes}) and (\ref{incompress}) which, for an
unbounded liquid, is given by the Oseen tensor,
\begin{equation}
  B^{21}_{ij}(\vecr) \simeq \frac{1}{8\pi\eta r} \left( \delta_{ij}
  + \frac{r_ir_j}{r^2} \right).
\label{Oseen3D}
\end{equation}
This leads, according to eq.\ (\ref{Bc}), to
\begin{equation}
\begin{split}
  &B^{\rm c}_{\rm L}(r\gg a) \simeq \frac{1}{4\pi\eta r},\\
  &B^{\rm c}_{\rm T1}(r\gg a) = B^{\rm c}_{\rm T2}(r\gg a) \simeq
  \frac{1}{8\pi\eta r}.
\end{split}
\label{Bc3D}
\end{equation}

Equations (\ref{Oseen3D}) and (\ref{Bc3D}) are independent of the
sizes and shapes of the particles. This universality is related to the
fact that they can be obtained, up to a numerical prefactor, solely
from conservation arguments. The force monopole associated with
particle 1 introduces a momentum source in the liquid. To conserve the
total momentum flux emanating from the source and passing through an
envelope of radius $r$, the local flux must decay as $1/r^2$. That
momentum flux is the liquid stress tensor, $\vecsigma\sim 1/r^2$,
whose shear part is related to liquid velocity as
$\vecsigma\sim\eta\nabla\vecv$; hence, $\vecv\sim 1/(\eta
r)$. Momentum conservation in 3D, therefore, dictates the following
form of the coupling mobility tensor: $B^{21}_{ij}\sim (\eta r)^{-1}
(\delta_{ij} + Cr_ir_j/r^2)$. The constant $C$ is then forced by mass
conservation (incompressibility), $\partial_i B^{21}_{ij}=0$, to be
$C=1$.

This argument remains intact when the two test particles are
surrounded by other particles, so long as the entire particulate
liquid conserves momentum (\ie remains translation-invariant). The
only thing that can change at sufficiently large inter-particle
distances is the prefactor in eqs.\ (\ref{Oseen3D}) and
(\ref{Bc3D}). The modified prefactor, depending on the volume fraction
$\phi$ of particles, defines an effective viscosity, $\eta_{\rm
eff}(\phi)$. An explicit calculation for an unbounded suspension of
hard spheres, to linear order in $\phi$, confirms this statement
\cite{ijc07}, yielding $\eta_{\rm eff}\simeq\eta(1+5\phi/2)$, in
agreement with Einstein's classical result \cite{Einstein}.

As particle 1 moves through the liquid, it perturbs not only the
liquid momentum density but also its mass density. (In the limit of an
incompressible liquid this perturbation does not disappear but is
accounted for by $p(\vecr)$, which is determined from the
incompressibility constraint.) To leading order in $a/r$ the mass
perturbation may be replaced by a mass dipole (a source and a sink). A
mass source would create a flow velocity proportional to $1/r^2$;
hence, the mass dipole creates a flow that decays as $1/r^3$. This
effect is manifest in the exact expression for the flow due to a
single translating rigid sphere \cite{Happel} and, consequently,
contributes to the Rotne--Prager mobility tensor
\cite{RotnePrager}, which is widely used in computer simulations of
suspensions and polymer solutions. The resulting correction to the
coupling mobility is smaller by an order of $(a/r)^2$ than the leading
$1/r$ term and is negligible, therefore, in the limit $r\gg a$.  The
dominant momentum-source contribution leads to a flow field of
monopolar shape and to strictly positive coupling coefficients [eq.\
(\ref{Bc3D})].

These consequences of the conservation of transverse momentum and mass
for the pair hydrodynamic interaction should be borne in mind as we
turn in the following sections to the more complicated cases of
confined liquids.

\section{Quasi-Two-Dimensional Systems}
%======================================
\label{sec_q2D}

In q2D systems one of the dimensions of the confined liquid (in the
$\zhat$ direction, say) is much smaller than the other two. Assuming
that this small width $w$ is not much larger than the particle size,
we ignore particle motion in the $\zhat$ direction. Thus, particles
move essentially in two dimensions, whereas the full dynamics of the
system (\eg momentum transport) remain three-dimensional.

Since the surface area of such a system scales with its volume, the
type of contact between the confined liquid and the environment plays
a crucial role and strongly affects the hydrodynamic interaction
between embedded particles. In \S\ref{sec_plates} and \S\ref{sec_soap}
we treat two useful limits for this contact---one in which transverse
momentum is completely absorbed by the boundaries, and another in
which it is fully conserved (corresponding, respectively, to no-slip
and slip boundary conditions for the liquid velocity). Section
\ref{sec_membrane} describes fluid membranes, which represent an
interesting intermediate between those two limiting behaviors.

\subsection{Confinement between two rigid surfaces}
%--------------------------------------------------
\label{sec_plates}

Hydrodynamic interactions in colloid suspensions confined between two
parallel solid plates have been thoroughly investigated, both
experimentally and theoretically, in the past several years
\cite{prl04,jpcm052,Santana,Jerzy,Alvarez,Rice,Davit}.  The plates, fixed
in the lab frame, break the translational symmetry of the
confined liquid; hence, liquid momentum is not conserved over
distances larger than $w$. The loss of transverse momentum is usually
taken into account by imposing no-slip boundary conditions, $\vecv=0$,
at the confining surfaces, leading to a finite momentum flux,
$\partial_z\vecv\neq 0$, into the plates. Consequently, the component
of the flow due to particle 1 which is analogous to eq.\
(\ref{Oseen3D}), \ie arising from momentum conservation, is
exponentially small in $r/w$.

Liquid mass, however, remains conserved. Hence, the mass-displacement
term introduced by particle 1, as described in \S\ref{sec_3D}, should
become the dominant contribution to the flow at $r>w$. Furthermore,
for $r\gg w$ the flow becomes essentially two-dimensional,
$\vecv(\vecr)$ lying in the $xy$ plane and having a symmetric
(parabolic) profile in the $\zhat$ direction. Thus, at a large
distance the flow due to the forced particle 1 looks as if it were due
to a 2D mass dipole. Such a flow decays as $1/r^2$ (since that of a
mass source in 2D decays as $1/r$ to preserve the flux through the
perimeter of an envelope of radius $r$). Consequently, the coupling
mobility tensor must be proportional to
$-r^{-2}(\delta_{ij}+Cr_ir_j/r^2)$, where the minus sign stems from
the direction of the mass dipole. (The pressure is higher in front of
the particle and lower behind it.)  Mass conservation,
$\partial_iB^{21}_{ij}=0$ ($i,j=x,y$), sets $C=-2$.  In addition, on
dimensional grounds, the dipole strength must be equal to $\alpha
w/\eta$, where $\alpha(a/w)$ is a dimensionless prefactor depending on
the confinement ratio $a/w$. The coupling tensor is, therefore,
\begin{equation}
  B^{21}_{ij}(\vecr) \simeq -\alpha(a/w) \frac{w}{\eta r^2} 
  \left( \delta_{ij} - 2\frac{r_ir_j}{r^2} \right),
\label{Oseen2plate}
\end{equation}
leading, according to eq.\ (\ref{Bc}), to
\begin{equation}
  B^{\rm c}_{\rm L,T}(r\gg w) \simeq \pm\alpha(a/w)
  \frac{w}{\eta r^2},
\label{Bc2plate}
\end{equation}
where the positive (negative) sign corresponds to the longitudinal
(transverse) interaction.

Identical results to eqs.\ (\ref{Oseen2plate}) and (\ref{Bc2plate})
can be obtained by considering the exact solution for the flow due to
a point force in this geometry \cite{Mochon}, applying a lubrication
approximation to eqs.\ (\ref{Stokes}) and (\ref{incompress})
\cite{Tsvi}, or treating the particles as suspended in a 2D Brinkman
fluid \cite{Brinkman}---\ie a fluid satisfying eqs.\ (\ref{Stokes})
and (\ref{incompress}) in 2D with an additional friction term $\sim
-(\eta/w^2)\vecv$ on the left-hand side of eq.\ (\ref{Stokes}). This
merely highlights the generality of the results, which arise from
conserved liquid mass in 2D and unconserved momentum. For example, it
is evident from the aforementioned arguments that allowing for finite
slip at the confining surfaces will not qualitatively change eqs.\
(\ref{Oseen2plate}) and (\ref{Bc2plate})---the fact that only part of
the transverse momentum imparted to the plates is lost does not change
the basic behavior of the q2D suspension as momentum-leaking and
mass-conserving. Nor will these asymptotic results for $r\gg w$ change
if we include the effect of particle motion in the third ($\zhat$)
dimension, as both the momentum monopole and mass dipole created by
such a motion will result in a flow which is exponentially small in
$r/w$ \cite{Mochon}. Consequently, even in cases of weak confinement,
$w\gg a$, where there may be many layers of particles between the two
surfaces, the crossover to the 3D hydrodynamic coupling of eq.\
(\ref{Bc3D}) will occur only at sufficiently small distances, $a\ll
r\ll w$. (In such cases the prefactor $\alpha$ will depend
on particle concentration.)

Thus, confinement between two rigid surfaces qualitatively changes the
pair hydrodynamic interaction. It strongly suppresses the
momentum-monopole contribution (from a long-ranged $1/r$ effect to an
exponential decay), while amplifying the mass-dipole one (from a 3D
$1/r^3$ effect to a 2D $1/r^2$ one). The amplification of the mass
term, in fact, makes the hydrodynamic interaction in this geometry
decay more slowly than the one near a single rigid surface (which
decays only as $1/r^3$) \cite{1wall1,1wall2}. The confinement also
changes the sign of the transverse interaction from positive [eq.\
(\ref{Bc3D})] to negative [eq.\ (\ref{Bc2plate})], which is a
consequence of the dipolar shape of eq.\ (\ref{Oseen2plate}). Most
interestingly, since the mass dipole induced by particle 1, $\alpha
w/\eta$, is unaffected by the presence of surrounding particles (so
long as the suspension of particles is sufficiently confined and/or
dilute so as not to have a correlation length smaller than $w$)
\cite{ijc07}, the hydrodynamic interaction is independent of particle
concentration. Thus, unlike the unconfined case of \S\ref{sec_3D}, the
effective viscosity, as defined by the prefactors of eq.\
(\ref{Oseen2plate}), is not modified. This statement has been verified
by an explicit calculation to first order in the particle area
fraction $\phi$, yielding a vanishing concentration correction to the
interaction at large distances \cite{jpcm052}. The leading correction
at high area fractions is a short-ranged effect reflecting the
equilibrium structure (pair correlation function) of the concentrated
suspension \cite{jpcm052}.

All the aforementioned predictions arising from eq.\
(\ref{Bc2plate})---the $1/r^2$ decay, the opposite-sign couplings, the
concentration-independence of the large-distance interaction---have
been confirmed to high accuracy in video-microscopy experiments
\cite{prl04,jpcm052,Santana}. 
In addition, the prefactor $\alpha$ is found to have a moderate
dependence on the confinement ratio. In the limit $a/w\rightarrow 0$
(yet continuing to assume that the particles lie at the mid-plane
between the two plates), it is analytically found as
$\alpha=3/(32\pi)\simeq 0.030$
\cite{Mochon}. The measured value for $a/w\simeq 0.45$ (quite
close to the upper bound of $1/2$) is $\alpha\simeq 0.019$
\cite{jpcm052}.

The dipolar hydrodynamic interactions of eq.\ (\ref{Bc2plate}) have
further fundamental consequences when the inversion symmetry in the
$xy$ plane is broken as well---\eg under driven flow in a microfluidic
channel, where the particles (or droplets) exhibit a new type of
density waves with an unusual dispersion relation \cite{Tsevi1}, or
when the confining plates are moved relative to one another
\cite{Davit}.

\subsection{Free-standing liquid films}
%--------------------------------------
\label{sec_soap}

The situation is drastically different when the q2D system is bounded
by vacuum or gas, as in a soap film. In this case both momentum and
mass are conserved, and the main effect of confinement is to make the
flow at $r\gg w$ essentially two-dimensional
\cite{Frydel1,Frydel2}. The momentum flux emanating from a momentum
monopole in 2D must decay as $1/r$ and, therefore, the flow velocity
due to the forced particle falls off logarithmically with
distance. This necessitates a cutoff length, $\kappa^{-1}$, which may
arise from the lateral system size, liquid inertia, or viscosity of
the outer medium, depending on the particular system. (In micron-scale
free-standing films it is the system size which usually determines
$\kappa$ \cite{soap1}.) The mass-dipole effect is of order
$(w/r)^2$ smaller and can be neglected.

The logarithmic decay slightly complicates the general procedure used
in \S\ref{sec_3D} and \S\ref{sec_plates} to deduce $\vecB^{21}$. The
coupling tensor is bound to be proportional to $[(C_1+C_2\ln\kappa
r)\delta_{ij} + (1+C_3\ln\kappa r)r_ir_j/r^2]$ ($i,j=x,y$). Mass
conservation in 2D, $\partial_iB^{21}_{ij}=0$ for any $r$, sets
$C_3=0$ and $C_2=-1$. The value of the last constant, $C_1$, depends
on the boundary condition imposed at the cutoff perimeter
$r=\kappa^{-1}$. For example, imposing a vanishing radial velocity at
the edge, $B^{21}_{ij}r_j|_{r=\kappa^{-1}}=0$, leads to $C_1=-1$. If
the strength of the 3D momentum monopole, associated with the forced
particle 1, is taken as unity (\ie a unit point force), then the
strength of the resulting 2D monopole is equal to $w$. This
additional requirement sets the prefactor of $\vecB^{21}$ to be
$(4\pi\eta w)^{-1}$.  Thus, in summary, we find
\begin{equation}
  B^{21}_{ij}(\vecr)\simeq \frac{1}{4\pi\eta w} \left[
  -\left(1+\ln(\kappa r)\right)\delta_{ij} + \frac{r_ir_j}{r^2} \right],
\label{Oseensoap}
\end{equation}
leading, according to eq.\ (\ref{Bc}), to
\begin{equation}
\begin{split}
  B^{\rm c}_{\rm L}(r\gg w) &\simeq -\frac{1}{4\pi\eta w}\ln(\kappa r),\\
  B^{\rm c}_{\rm T}(r\gg w) &\simeq -\frac{1}{4\pi\eta w}
  \left[1+\ln(\kappa r)\right].
\end{split}
\label{Bcsoap}
\end{equation}
(The vanishing of the longitudinal interaction vs.\ the finite value
of the transverse one at the perimeter $r=\kappa^{-1}$ stem from the
specific boundary conditions imposed above.) The prefactor in eqs.\
(\ref{Oseensoap}) and (\ref{Bcsoap}) defines a two-dimensional film
viscosity, $\eta_{\rm m}=\eta w$. As in the unconfined case of
\S\ref{sec_3D}, and unlike the two-plate confinement of
\S\ref{sec_plates}, eqs.\ (\ref{Oseensoap}) and (\ref{Bcsoap}) are
independent of the particle size and shape. This is because they stem
directly from the unit force introduced by particle 1 and not from the
effective mass dipole associated with it.

The ultra-long-ranged hydrodynamic interaction described by eq.\
(\ref{Bcsoap}) has recently been observed in soap films containing
colloid particles \cite{soap1,soap2}. The crossover between this 2D
behavior and the 3D one for $r<w$ has been demonstrated as well
\cite{soap2}.

Since hydrodynamic correlations in the film are carried over large
distances by transverse momentum, we expect the 2D viscosity, entering
the prefactors of eqs.\ (\ref{Oseensoap}) and (\ref{Bcsoap}), to be
modified by the presence of surrounding particles, similar to the
unconfined case. Embedding rigid cylindrical inclusions of height $w$
and radius $a$ in the film makes the problem purely 2D. An explicit
calculation of the modified prefactors, to linear order in the
particle area fraction $\phi$, leads to $\eta_{\rm m,eff}=\eta_{\rm
m}(1+2\phi)$ \cite{bpj09}, in agreement with the known effective
viscosity of such a 2D suspension of hard disks \cite{Belzons}.

\subsection{Membranes}
%---------------------
\label{sec_membrane}

Fluid membranes provide a particularly important example of
hydrodynamic interactions in confined geometry. Membranes form the
envelopes of all living cells and are also used to make vesicles
(liposomes) for various applications. The membrane is a self-assembled
bilayer of amphiphilic molecules (\eg lipids), which are free to move
in the lateral directions, thus forming a q2D liquid. Biomembranes
contain also a high concentration of membrane-embedded proteins, whose
motion is confined to the membrane surface as well. Since the membrane
viscosity $\eta$ is much higher than the viscosity $\eta_{\rm f}$ of
the surrounding aqueous medium, yet not infinitely so (typically by a
factor of $\sim 10^3$), this q2D system presents an interesting
intermediate between the two systems studied in \S\ref{sec_plates} and
\S\ref{sec_soap}.  From the conservation arguments that underlie our
discussion it is clear that one should distinguish between two
cases---one in which the membrane is freely suspended in the solution,
and another in which it is immobilized.

In the first case the system is translationally invariant and, hence,
conserves momentum. The hydrodynamics of such a membrane was first
studied by Saffman and Delbr\"uck \cite{Saffman,SD} and later, using a
different approach, by Levine and MacKintosh \cite{Levine}. The large
viscosity contrast introduces a length scale, $\kappa^{-1}=\eta
w/(2\eta_{\rm f})$, which is much larger than the molecular thickness
of the membrane and protein size, $\kappa^{-1}\gg w\sim a$. The
Saffman-Delbr\"uck length $\kappa^{-1}$, which is typically of micron
scale, sets the distance beyond which momentum is transported
primarily through the surrounding liquid rather than through the
membrane.

At distances $a\ll r\ll\kappa^{-1}$ the membrane behaves much like the
soap film of \S\ref{sec_soap}, conserving both momentum and mass in
2D. Over such intermediate distances, therefore, the coupling mobility
tensor of a protein pair should be [cf.\ eq.\ (\ref{Oseensoap})]
\begin{equation}
  B^{21}_{ij}(\vecr)\simeq \frac{1}{4\pi\eta_{\rm m}} \left[
  -\left(1+\ln(\kappa' r)\right)\delta_{ij} + \frac{r_ir_j}{r^2} \right],
\label{Oseenmemb1}
\end{equation}
where $\eta_{\rm m}=\eta w$ is the membrane 2D viscosity, and
$\kappa'\sim\kappa$ up to a numerical constant. This leads to [cf.\
eq.\ (\ref{Bcsoap})]
\begin{equation}
\begin{split}
  B^{\rm c}_{\rm L}(a\ll r\ll \kappa^{-1}) &\simeq
  -\frac{1}{4\pi\eta_{\rm m}}\ln(\kappa' r),\\ 
  B^{\rm c}_{\rm T}(a\ll r\ll\kappa^{-1}) 
  &\simeq -\frac{1}{4\pi\eta_{\rm m}} \left[1+\ln(\kappa' r)\right].
\end{split}
\label{Bcmemb1}
\end{equation}

At sufficiently large distances, $r>\kappa^{-1}$, momentum is
transported through the surrounding liquid as well and is conserved in
3D rather than within the q2D membrane. As in \S\ref{sec_3D}, this
dictates the following form for the coupling mobility tensor:
$B^{21}_{ij}\sim (\eta_{\rm f} r)^{-1} (C\delta_{ij} +
r_ir_j/r^2)$. Yet, unlike the unconfined case of \S\ref{sec_3D}, the
membrane (lipid) mass is conserved in 2D, $\partial_i B^{21}_{ij}=0$
($i,j=x,y$), which sets $C=0$. We thus have
\begin{equation}
  B^{21}_{ij}(\vecr) \simeq \frac{\epsilon}{\eta_{\rm f}} \frac{r_ir_j}{r^3},
\label{Oseenmemb2}
\end{equation}
where $\epsilon$ is a dimensionless prefactor. This leads, according to 
eq.\ (\ref{Bc}), to 
\begin{equation}
  B^{\rm c}_{\rm L}(r\gg\kappa^{-1}) \simeq \frac{\epsilon}{\eta_{\rm f}r},
\label{Bcmemb2}
\end{equation}
and a transverse interaction $B^{\rm c}_{\rm T}$ which is of order
$(\kappa r)^{-1}$ smaller. Equations (\ref{Oseenmemb2}) and
(\ref{Bcmemb2}) are independent of any membrane property, reflecting
the fact that the hydrodynamic interaction over such long distances is
mediated solely by the surrounding medium. In particular, the
interaction in this regime will not be modified by the presence of
other membrane inclusions, \ie it is independent of protein
concentration.

A more detailed calculation \cite{bpj09,Levine} yields the
crossover between these two distance regimes, as well as the values
for the numerical factors left unspecified above:
$\kappa'=(e^{\gamma_{\rm E}-1/2}/2)\kappa$ ($\gamma_{\rm E}$ being the
Euler constant), and $\epsilon=1/(4\pi)$. An explicit calculation of
the effect of rigid cylindrical inclusions on the interaction
\cite{bpj09} shows that, to linear order in the area fraction $\phi$
of inclusions, one can replace $\eta_{\rm m}$ with $\eta_{\rm
m,eff}=\eta_{\rm m}(1+2\phi)$, as in \S\ref{sec_soap}, except that
this substitution should be made also in the Saffman-Delbr\"uck
length, $\kappa^{-1}=\eta_{\rm m}/(2\eta_{\rm f})\rightarrow \eta_{\rm
m,eff}/(2\eta_{\rm f})$.

In the second relevant case, where the membrane is immobilized (\eg
supported on solid substrates) the system is not translationally
invariant and, hence, does not conserve momentum. Another length
scale, $\lambda$, should be considered, beyond which membrane momentum
is lost to the solid. The Saffman-Delbr\"uck length being 
of micron scale, we have in general $\lambda\ll\kappa^{-1}$. Since the
membrane conserves mass in 2D, this scenario is equivalent to the one
considered in \S\ref{sec_plates}, with $\lambda$ replacing $w$, and
the particles can be treated as embedded in a 2D Brinkman fluid
\cite{ES,Suzuki,Komura}.

The dependencies summarized in eqs.\
(\ref{Oseenmemb1})--(\ref{Bcmemb2}) were observed in the dynamics of
domains \cite{McConnell,Keller} and colloids \cite{Weeks2} embedded in
monolayers of amphiphilic molecules at the water--air interface.  (In
such systems, where one of the bounding fluids has a vanishingly small
viscosity, one should replace $\eta_{\rm f}$ in the aforementioned
results with $\eta_{\rm f}/2$.)  Two-particle tracking experiments for
membrane-embedded proteins, to our best knowledge, have not been
performed yet. Nor are we aware of similar experiments involving
immobilized membranes.

\section{Quasi-One-Dimensional Systems}
%======================================
\label{sec_q1D}

Following the line of argument of \S\ref{sec_q2D}, a liquid confined
in a linear channel with rigid walls does not conserve momentum beyond
a distance comparable to the channel width $w$. The flow at a distance
$x$ from the momentum monopole due to the forced particle 1,
therefore, is exponentially small in $x/w$. However, unlike the q2D
channel, the flow due to the mass dipole is also short-ranged. This is
because at $x\gg w$ the flow velocity becomes essentially
one-dimensional, pointing in the $\xhat$ direction. Since a mass
source in 1D creates a flow which is uniform in $x$, a 1D mass dipole,
as well as all higher moments, create no flow at all. Thus, q1D
confinement in a rigid channel leads to exponential screening of the
hydrodynamic interaction beyond $x>w$. This has been confirmed
experimentally \cite{prl02}.  The effect of the other particles in the
q1D suspension on this screened interaction has been accounted for as
well \cite{Xu}.

As in \S\ref{sec_plates}, introducing partial slip at the channel
boundaries should not change this qualitative result. Nevertheless, if
the particles are made bigger so as to fit the channel cross-section,
one expects the flows to become plug-like, and the range of coupling
between particles in an incompressible liquid to tend to
infinity. Indications for such a divergence were observed in the
dynamics of droplets driven in narrow microfluidic channels
\cite{Tsevi2}.

\section{Gels and Porous Media}
%==============================
\label{sec_porous}

A commonly encountered confinement is that of a liquid pervading a
solid matrix, such as a porous medium or a polymer network. It is
often (wrongly) assumed that the hydrodynamic interaction between
particles embedded in such a system is screened beyond a distance
comparable to the correlation length $\xi$ of the matrix. The liquid,
whose translational symmetry is broken by the stationary matrix, does
not conserve momentum, and the flow at a distance $r$ away from the
momentum monopole due to particle 1 will be exponentially small in
$r/\xi$. Liquid mass, however, is conserved. Thus, at a large distance
the flow due to the forced particle 1 looks as if it were due to a 3D
mass dipole \cite{ijc07}. Such a flow decays as $1/r^3$. Consequently,
the coupling mobility tensor must be proportional to
$-r^{-3}(\delta_{ij}+Cr_ir_j/r^2)$, where the minus sign is implied by
the mass dipole direction. Mass conservation in 3D,
$\partial_iB^{21}_{ij}=0$ ($i,j=x,y,z$), sets $C=-3$.  Additionally,
on dimensional grounds, the dipole strength must be equal to $\gamma
\xi^2/\eta$, where $\gamma(a/\xi)$ is a dimensionless prefactor
depending on the confinement ratio $a/\xi$. The coupling tensor is,
therefore,
\begin{equation}
  B^{21}_{ij}(\vecr) \simeq -\gamma(a/\xi) \frac{\xi^2}{\eta r^3} 
  \left( \delta_{ij} - 3\frac{r_ir_j}{r^2} \right),
\label{Oseengel}
\end{equation}
leading, according to eq.\ (\ref{Bc}), to
\begin{equation}
\begin{split}
  B^{\rm c}_{\rm L}(r\gg\xi) &\simeq 2\gamma(a/\xi) \frac{\xi^2}{\eta r^3},\\
  B^{\rm c}_{\rm T1}(r\gg\xi) &= B^{\rm c}_{\rm T2}(r\gg\xi) 
  \simeq -\gamma(a/\xi) \frac{\xi^2}{\eta r^3}.
\end{split}
\label{Bcgel}
\end{equation}
Identical results to eqs.\ (\ref{Oseengel}) and (\ref{Bcgel}) can be
obtained by considering the particles as embedded in a 3D Brinkman
fluid \cite{Brinkman}---\ie a fluid that satisfies eqs.\
(\ref{Stokes}) and (\ref{incompress}) in 3D with an additional
friction term $\sim -(\eta/\xi^2)\vecv$ on the left-hand side of eq.\
(\ref{Stokes}) \cite{LongAjdari}.

These results for confinement in a solid matrix are similar to those of
\S\ref{sec_plates}, and so are their consequences. (Indeed, the two-plate 
geometry may be viewed as a particular example of a porous medium.)
Confinement qualitatively changes the hydrodynamic interaction by
strongly suppressing the momentum-monopole contribution (from a
long-ranged $1/r$ effect to an exponential decay), while keeping the
mass-dipole one intact. The resulting $1/r^3$ interaction decays
faster with distance than the unconfined one but is still long-ranged.
The confinement also changes the sign of the transverse interaction
from positive [eq.\ (\ref{Bc3D})] to negative [eq.\ (\ref{Bcgel})],
which is a consequence of the dipolar shape of eq.\
(\ref{Oseengel}). As in \S\ref{sec_plates}, we reach the surprising
conclusion that, so long as the suspension of particles is
sufficiently confined and dilute so as not to have a correlation
length smaller than $\xi$, the hydrodynamic interaction between
particles embedded in the matrix is independent of particle
concentration.  This is because the mass dipole induced by particle 1,
$\gamma \xi^2/\eta$, will be unaffected by the presence of surrounding
particles. In this case the prefactors in eqs.\ (\ref{Oseengel}) and
(\ref{Bcgel}) are independent of concentration, and the effective
viscosity of the suspension, as defined by those prefactors, is
$\eta_{\rm eff}(\phi)=\eta$.

To our best knowledge, the aforementioned effects have never been
experimentally observed. They should be readily testable using
two-particle tracking, \eg in a polymer gel. For eqs.\
(\ref{Oseengel}) and (\ref{Bcgel}) to hold in such an experiment, the
particles should be sufficiently small, $a<\xi$, and the frequency
sufficiently low, so as to avoid viscoelastic effects.

\section{Discussion}
%===================
\label{sec_discuss}

The scenarios addressed in the preceding sections demonstrate how the
hydrodynamic interaction between particles embedded in a confined
fluid is strongly and differently affected by the confinement,
depending on the specific geometry.  On the one hand, confinement may
suppress the transverse-momentum contribution to the interaction,
which is the dominant one in unconfined fluids. This occurs when solid
boundaries break the translational symmetry of the confined fluid. On
the other hand, the mass-displacement contribution, which is
negligible in unconfined fluids, may be either amplified, due to
a reduction of the effective dimensionality of the flow, or left
intact. Additionally, even severe confinement may have a negligible
effect when the response of the outer medium, rather than that of the
confined fluid, is dominant (as is the case for the large-distance
interaction in membranes). As a result of these opposing trends, and
somewhat against one's naive expectation, in most cases confinement
does {\em not} lead to overall suppression, or screening, of the
hydrodynamic interaction at distances larger than the confinement
width $w$. For instance, the hydrodynamic interactions in fluid
membranes, a two-plate geometry, and porous matrices decay only
algebraically, as $1/r$, $1/r^2$, and $1/r^3$, respectively.

Several of the theoretical results presented here have been
convincingly confirmed in experiment (\eg the ones pertaining to the
q2D geometries of \S\ref{sec_plates} and \S\ref{sec_soap}). Yet, other
predictions are still to be tested---primarily those related to gels
and liquid-filled porous media (\S\ref{sec_porous}). The dipolar shape of
the coupling in this case (as well as in the two-plate geometry)
yields a vanishing effect upon angular averaging
\cite{Tsvi}. Consequently, the long-ranged interaction will not be
observed in conventional scattering measurements. Two-particle
tracking, nonetheless, should readily reveal it.

The agreement between theory and available experimental results
suggests that the combination of assumptions put forth in
\S\ref{sec_3D}, on which the entire analysis has relied, is valid
under common experimental conditions. It would be beneficial,
nevertheless, to examine the consequences of relaxing some of these
assumptions under different conditions. The following key issues are
left open for future study.
(i) As has been mentioned above, introducing partial liquid slip at the
rigid boundaries should not lead to a qualitative change in the
results. However, considering finite rigidity of the outer medium will
allow momentum exchange with it, thus affecting the hydrodynamic
interaction at large distances.
(ii) The limit of zero Reynolds number has allowed us to ignore the
time dependence of the hydrodynamic interaction. It has been already
shown that the switch of dominance between momentum and mass transport
in certain confined geometries should lead to a drastically faster
buildup of the interaction \cite{ijc07}. Including temporal dependence
is obviously crucial also when considering viscoelastic effects.
(iii) The assumption of liquid incompressibility implies that sound
propagates infinitely fast through the system. Confinement by rigid
surfaces is known to strongly affect the sound modes of the confined
liquid, making them diffusive \cite{Frenkel,Frydel2}. The resulting
diffusivity, $c^2w^2/\nu$, is still very large ($\sim 1$ m$^2$/s for
water in a micron-scale channel), yet the qualitatively different
sound propagation may have an interesting effect on the short-time
hydrodynamic coupling.
(iv) Breaking inversion symmetry by driven flow has already been
shown, both experimentally and theoretically, to lead to novel
dynamics of confined particles \cite{Tsevi1,Tsevi2}. These new
findings clearly call for further investigation---\eg of confined
sedimentation.

This article has shown how key results concerning the hydrodynamic
interaction in varied confined geometries can be simply and accurately
derived from the effect of confinement on the conserved fields of the
fluid.  Owing to their generality, such arguments may be found helpful
in other scenarios. They can provide asymptotes against which more
detailed theories or simulations are to be tested.  Furthermore,
analogous approaches can be applied to confinement effects on other
medium-induced interactions. For example, when particles are confined
between two plates of high dielectric permittivity, their static
(London) dispersion interaction is suppressed due to the concentration
of field lines in the plates, whereas the dynamic (Casimir-Polder) one
is enhanced due to the reduced effective dimensionality for radiation
\cite{prl05}.

\section*{Acknowledgment}

This research has been supported by the Israel Science Foundation (Grant no.\ 588/06).

% References
%===========

\end{document}